\documentclass[aps,showpacs,superscriptadress,preprint]{revtex4}

\begin{document}
\title{The intrinsic degree of freedom for quasiparticle in thermodynamics with medium effects}

\author{Shaoyu Yin$^1$ and Ru-Keng
Su$^{1,2}$\footnote{rksu@fudan.ac.cn}} \affiliation{
\small 1. Department of Physics, Fudan University,Shanghai 200433, P. R. China\\
\small 2. CCAST(World Laboratory), P.O.Box 8730, Beijing 100080, P. R. China\\
}

\begin{abstract}
The thermodynamics with medium effects expressed by the temperature-
and density-dependent effective mass of quasiparticle is studied.
Series difficulties and many wrangles in references due to the
extraordinary parameter dependence are addressed. A new independent
intrinsic degree of freedom of quasiparticle $m^*$ in the equation
of reversible process is introduced to clear the ambiguity. We prove
all results are self-consistent.
\end{abstract}


\maketitle

It is generally accepted that the effective mass of particle will
change with temperature and density due to its interaction with the
environment, usually referred as the medium effects. This phenomenon
is not only demonstrated by theoretical studies, for example,
Brown-Rho scaling \cite{Brown:1991}, Quantum Chromodynamics(QCD) sum
rules \cite{Hatsuda:1993}, vacuum polarization Feynman diagrams
calculated by Thermo-Field Dynamics \cite{Zhang:1997,Shiomi:1994},
\textit{etc.}, but also by experiment \cite{Lolos:1998}.

Although the density- and temperature-dependent quasiparticle mass
$m^*(T,\rho)$ is commonly used to mimic the medium effects,
difficulty emerges when we discuss the thermodynamics of the system
with such quasiparticle. In thermodynamics, a proper choice of
independent variables will have a suitable characteristic
thermodynamical function, from which all the thermodynamic
quantities can be obtained by partial derivatives without
integration. For example, with variables temperature $T$, volume $V$
and chemical potential $\mu$, the characteristic function is the
thermodynamical potential $\Omega=\Omega(T,V,\mu)$. From the exact
differential relation
\begin{equation}
d\Omega=-SdT-pdV-Nd\mu,
\end{equation}
we have
\begin{equation}
S=-(\frac{\partial \Omega}{\partial T})_{V,\mu},\qquad
p=-(\frac{\partial \Omega}{\partial V})_{T,\mu},\qquad
N=-(\frac{\partial \Omega}{\partial \mu})_{T,V},
\end{equation}
where $S$, $p$ and $N$ are entropy, pressure and particle number
respectively. Other thermodynamic quantities such as internal energy
$U$, Helmholtz free energy $F$, enthalpy $H$, Gibbs function $G$,
\textit{etc.}, can also be calculated by the combination of the
quantities we obtained, based on their definitions or relations.

But for the case with effective mass, the relativistic dispersion
relation for a quasiparticle with energy $\epsilon$ and momentum $k$
becomes
\begin{equation}
\epsilon=(k^2+m^*(T,\rho)^2)^{1/2},
\end{equation}
so $\Omega$ is not only a function of $T$, $V$, $\mu$, but also
depends explicitly on the quasiparticle mass,
$\Omega=\Omega(T,V,\mu,m^*(T,\rho))$. How to tackle the
thermodynamics self-consistently is still a problem. There have been
many wrangles in present references \cite{Chakrabarty:1993,
Benrenuto:1995dual,Peng:1999,Peng:2000,Zhang:2002,Wen:2005,Wang:2000}.
The difficulty comes from the first and second laws of reversible
process thermodynamics expressed in Eq.(1) and the partial
derivatives as in Eq.(2). Obviously, some extra terms involving the
derivatives of $m^*$ will emerge. Unfortunately, these extra terms
for different treatments in different references contradict each
other. For example, for ideal gas system of quasiparticle with
effective mass $m^*=m^*(\rho)$, the pressure and energy density
$\varepsilon$ were given by
\begin{eqnarray}
p&=&-\widetilde{\Omega}\equiv-\frac{\Omega}{V},\\
\varepsilon\equiv\frac{U}{V}&=&\widetilde{\Omega}+\sum_{i}\mu_{i}\rho_{i}-
T\frac{\partial\widetilde{\Omega}}{\partial T},
\end{eqnarray}
in Ref.\cite{Chakrabarty:1993}; given by
\begin{eqnarray}
p&=&-(\frac{\partial(\widetilde{\Omega}/\rho)}{\partial(1/\rho)})_{T,\{\mu_i\}}
=-\widetilde{\Omega}+\rho(\frac{\partial\widetilde{\Omega}}{\partial
\rho})_{T,\{\mu_i\}}\\
\varepsilon&=&\widetilde{\Omega}-\rho(\frac{\partial\widetilde{\Omega}}{\partial
\rho})_{T,\{\mu_i\}}+\sum_i\mu_i\rho_i-T(\frac{\partial\widetilde{\Omega}}{\partial
T})_{\{\mu_i\},\rho}.
\end{eqnarray}
in Ref.\cite{Benrenuto:1995dual}; and given by
\begin{eqnarray}
p&=&-\widetilde{\Omega}+\rho(\frac{\partial\widetilde{\Omega}}{\partial
\rho})_{T,\{\mu_{i}\}},\\
\varepsilon&=&\widetilde{\Omega}+\sum_{i}\mu_{i}\rho_{i}-T(\frac{\partial\Omega}{\partial
T})_{\{\mu_{i}\},\rho},
\end{eqnarray}
in Ref.\cite{Peng:2000}. For $m^*=m^*(T,\rho)$, $p$ and
$\varepsilon$ read
\begin{equation}
p=-\widetilde{\Omega}-V\frac{\partial\widetilde{\Omega}}{\partial
V}+\rho\sum_{i} \frac{\partial\widetilde{\Omega}}{\partial
m_{i}}\frac{\partial m_{i}}{\partial \rho},
\end{equation}
\begin{equation}
\varepsilon=\widetilde{\Omega}-\sum_{i}\mu_{i}
\frac{\partial\widetilde{\Omega}}{\partial\mu_{i}}-
T\frac{\partial\widetilde{\Omega}}{\partial
T}-T\sum_{i}\frac{\partial\widetilde{\Omega}} {\partial
m_{i}}\frac{\partial m_{i}}{\partial T},
\end{equation}
in Ref.\cite{Wen:2005}. The ambiguity arises from the variable
$\rho$ in $m^*$, because it is not one of the characteristic
variables in $\Omega(T,V,\mu)$. Even in Ref.\cite{Wang:2000}, a
supplement term $\Omega_{\alpha}$ which satisfied
$\frac{\partial(\widetilde{\Omega}+\Omega_{\alpha})}
{\partial\rho}=0$ was introduced to compensate the density
dependence of thermodynamical potential. The confusion in the market
tells us this problem of common interest is still unsolved.

This letter involves from an attempt to clear above ambiguity and
give a treatment to calculate the thermodynamic quantities from
partial derivatives self-consistently.

Noticing that at a fixed instant of reversible process, the system
is at an equilibrium state. Denote the temperature and density are
$T_0$ and $\rho_0$ respectively, then the effective mass of the
quasiparticle becomes constant $m^*(T_0,\rho_0)=m_0$. The system
reduces to a usual ideal gas system with constant mass $m_0$
quasiparticles. For this equilibrium state, the corresponding
thermodynamic quantities can be directly obtained. For example, for
the case of one component Fermi system:
\begin{eqnarray}
N&\equiv&\rho V=\sum_ig_in_i=\sum_i\frac{g_i}{e^{\beta(\sqrt{m_0^2+k_i^2}-\mu)}+1},\\
\Omega&=&-\sum_ig_ikT\ln(1+e^{-\beta(\sqrt{m_0^2+k_i^2}-\mu)}),\\
U&=&\sum_ig_in_i\epsilon_i=\sum_i\frac{g_i\sqrt{m_0^2+k_i^2}}{e^{\beta(\sqrt{m_0^2+k_i^2}-\mu)}+1},\\
G&=&N\mu=\sum_ig_in_i\mu=\sum_i\frac{g_i\mu}{e^{\beta(\sqrt{m_0^2+k_i^2}-\mu)}+1},\\
S&=&\frac{U-\Omega-G}{T},\\
p&=&-\frac{\Omega}{V},
\end{eqnarray}
where $n_i$ is the particle number of the $i$th state and $g_i$ is
the corresponding degeneracy.

We see, from Eqs.(12)-(17), the contribution of medium effect is
included in the effective value of mass, and appears in the
exponential of the Fermi distribution. A remarkable property of
these formulae is that the extra terms related to the partial
derivative of $m^*$ does not appear. This is reasonable because
these thermodynamic quantities are functions of state, they do not
depend on the change of the quasiparticle mass.

To compare above treatment with others given in
Ref.\cite{Chakrabarty:1993,Benrenuto:1995dual,Peng:1999,Peng:2000,
Zhang:2002,Wen:2005,Wang:2000}, as an example, we calculate the
entropy of the ideal quasiparticle system. Denote the entropy
calculated by equilibrium state as $S_{sta}$. From Eq.(13)-(16) we
have:
\begin{eqnarray}
S_{sta}&=&\frac{U-G-\Omega}{T}\nonumber\\
&=&\sum_ig_i[\frac{n_i\sqrt{m_0^2+k_i^2}-n_i\mu}{T}
+k\ln(1+e^{-\beta(\sqrt{m_0^2+k_i^2}-\mu)})]\nonumber\\
&=&k\sum_ig_i[n_i\beta(\sqrt{m_0^2+k_i^2}-\mu)+\ln(\frac{e^{\beta(\sqrt{m_0^2+k_i^2}-\mu)}+1}{e^{\beta(\sqrt{m_0^2+k_i^2}-\mu)}})]\nonumber\\
&=&k\sum_ig_i[n_i\ln(\frac{1}{n_i}-1)+\ln(\frac{1}{1-n_i})]\nonumber\\
&=&-k\sum_ig_i[n_i\ln(n_i)+(1-n_i)\ln(1-n_i)].
\end{eqnarray}
This is a familiar formula for equilibrium state whose physical
meaning is transparent. While we denote the entropy calculated by
the partial derivative of $\Omega$ following Eqs.(1) and (2) as
$S_{der}$, we have
\begin{eqnarray}
S_{der}&=&-(\frac{\partial\Omega(T,V,\mu,m^*(T,\rho))}{\partial
T})_{V,\mu}\nonumber\\
&=&-(\frac{\partial\Omega(T,V,\mu,m^*)}{\partial
T})_{V,\mu,m^*}-(\frac{\partial\Omega(T,V,\mu,m^*)}{\partial
m^*})_{T,V,\mu}(\frac{\partial m^*(T,\rho)}{\partial T})_{V,\mu},
\end{eqnarray}
and for the same equilibrium state with $m^*(T_0,\rho_0)=m_0$,
\begin{eqnarray}
&&-(\frac{\partial\Omega(T,V,\mu,m^*)}{\partial
T})_{V,\mu,m^*=m_0}\nonumber\\
&=&k\sum_ig_i\ln(1+e^{(\mu-\sqrt{m_0^2+k_i^2})/kT})+kT\sum_ig_i\frac{e^{(\mu-\sqrt{m_0^2+k_i^2})/kT}
(\frac{\sqrt{m_0^2+k_i^2}-\mu}{kT^2})}{1+e^{(\mu-\sqrt{m_0^2+k_i^2})/kT}}\nonumber\\
&=&k\sum_ig_i\ln(\frac{1}{1-n_{i}})+k\sum_ig_in_{i}\ln(\frac{1}{n_{i}}-1)\nonumber\\
&=&-k\sum_ig_i[n_{i}\ln n_{i}+(1-n_{i})\ln(1-n_{i})]=S_{sta}.
\end{eqnarray}

The first term of Eq.(19) is just the result given by equilibrium
state. The difference between these two treatments is significant.
They can not be accorded together. Noticing that the contribution of
medium effect at equilibrium state is included within the value of
$m^*$ in the distribution, and the entropy describing disorder of
quasiparticles in a system dose not depend on the intrinsic quantity
such as effective mass of quasiparticle, the correctness of
$S_{sta}$ is obvious.

Furthermore, if we calculate the particle number $N$ from Eqs.(1)
and (2), we obtain
\begin{equation}
N_{der}=N_{sta}-(\frac{\partial\Omega(T,V,\mu,m^*)}{\partial
m^*})_{T,V,\mu}(\frac{\partial m^*(T,\rho)}{\partial\mu})_{T,V},
\end{equation}
where $N_{sta}$ is calculated from the equilibrium state shown in
Eq.(12). Since $\rho=N/V$ is usually taken as a constraint to decide
the value of $\mu$, $(\frac{\partial
m^*(T,\rho)}{\partial\mu})_{T,V}$ dose not vanish if $m^*$ depends
on $\rho$ explicitly, so the extra term in Eq.(21) modifies the
particle number of the system. Since the number of quasiparticle
will not be changed by the medium effect, the result of Eq.(21) is
incorrect.

To get a self-consistent calculation between equilibrium state and
reversible process in thermodynamics, we are compelled by above
results to introduce an intrinsic degree of freedom $m^*$ for
quasiparticle in thermodynamic system with medium effect, and
rewrite Eq.(1) as:
\begin{equation}
d\Omega=-SdT-pdV-Nd\mu+Xdm^*,
\end{equation}
then Eq.(2) becomes
\begin{equation}
S=-(\frac{\partial \Omega}{\partial T})_{V,\mu,m^*},\quad
p=-(\frac{\partial \Omega}{\partial V})_{T,\mu,m^*},\quad
N=-(\frac{\partial \Omega}{\partial \mu})_{T,V,m^*},\quad
X=(\frac{\partial\Omega}{\partial m^*})_{T,V,\mu},
\end{equation}
where $X$ is an extensive quantity corresponding to the intensive
variable $m^*$. In Eq.(22), the intrinsic degree of freedom $m^*$
for quasiparticle has been added as an independent variable in
thermodynamic system. The quantities $S$, $p$ and $N$ shown in
Eq.(23) are in agreement with the results obtained by the formulae
of equilibrium state in Eqs.(12)-(17).

Usually the thermodynamic parameters such as $S$, $p$, $T$... depend
on the whole system. They are independent with the intrinsic
property of the particle or subsystem, no matter the subsystem is a
simple point particle or a quasiparticle with inner structure and
different intrinsic properties. Ordinary thermodynamic variables
depend on the collection of the subsystem only. Similarly, the mass
is an intrinsic quantity of a particle, it does not affect on
collective thermodynamic properties of the whole system. When we
consider the medium effect, and summarize this effect into the
effective mass $m^*(T,\rho)$ under quasiparticle approximation, the
dynamic interaction can be concentrated on the effective mass $m^*$
of quasiparticle by using the finite temperature quantum field
calculation \cite{Das:1997}. The thermodynamic variables cannot
describe these micro dynamic interactions. We must choose new
variables to represent these dynamic interactions or the medium
effect. Obviously, the effective mass $m^*$ appears as the suitable
independent variable. As we know that the number of independent
thermodynamic variables of a normal system obey the Gibbs' phase
rule \cite{Reichl:1980}, here we extend the Gibbs' phase rule to
include the medium effect by adding a new independent variable, the
intrinsic degree of freedom for quasiparticle, effective mass $m^*$.

In summary, we have introduced a new intrinsic degree of freedom
$m^*$ for the system of quasiparticle. We have proved that the
thermodynamic quantities calculated by the partial derivatives
concerning this independent variable are in agreement with those
obtained by the equilibrium state. The difficulties and
controversies in previous references are removed. Since the
effective mass of quasiparticle is quite widely used in many aspects
of physics, we hope our new statement may help the study on the
thermodynamic properties of such system.

This work is supported in part by NNSF of China.

\end{document}